# Optical Identification of Quasar 0917+7122 in the Direction of an Extreme-Ultraviolet Source

**Dan Maoz, Eran O. Ofek, Amotz Shemi**

School of Physics & Astronomy and Wise observatory
Tel-Aviv University, Tel-Aviv 69978, ISRAEL

**Aaron J. Barth, Alexei V. Filippenko**

Department of Astronomy
University of California, Berkeley, CA 94720, USA

**M. S. Brotherton, Beverley J. Wills, D. Wills**

McDonald Observatory and Astronomy Department
University of Texas, Austin, TX 78712, USA

**Felix J. Lockman**

National Radio Astronomy Observatory
P.O. Box 2, Green Bank, WV 24944-0002, USA

# Abstract

We report the optical identification of an $R = 18.3$ mag, $z = 2.432$ quasar at the position of a 6 cm radio source and a faint *ROSAT* PSPC X-ray source. The quasar lies within the error circles of unidentified extreme-UV (EUV) detections by the *EUVE* and *ROSAT WFC* all-sky surveys at $\sim 400$ Å and $\sim 150$ Å, respectively. A 21 cm H I emission measurement in the direction of the quasar with a $21'$-diameter beam yields a total H I column density of $N_H = 3.3 \times 10^{20}$ cm$^{-2}$, two orders of magnitude higher than the maximum allowed for transparency through the Galaxy in the EUV. The source of the EUV flux is therefore probably nearby ($\lesssim 100$ pc), and unrelated to the quasar.

# 1  Introduction

The *ROSAT Wide Field Camera (WFC)* and the *Extreme Ultraviolet Explorer (EUVE)* missions carried out all-sky surveys at extreme-ultraviolet (EUV) wavelengths in 1990-91 and 1992-93, respectively (Pounds et al. 1993; Malina et al. 1994; Bowyer et al. 1994). The *EUVE* all-sky survey was carried out in four broad bands, centered at approximately 100 Å, 200 Å, 400 Å, and 550 Å. The *ROSAT* WFC all-sky survey was conducted in two EUV bands, centered at about 100 Å and 150 Å, and similar to the 100 Å and 200 Å bands in the *EUVE* survey. The two experiments had comparable sensitivities in their common bands.

While the majority of the EUV sources detected by the two missions have been identified, mostly with white dwarfs and active late-type stars, about 15% of the EUV sources still have no identified optical counterparts. As part of an ongoing program of optical identification of EUV sources (Maoz, Ofek, & Shemi 1996) it was noticed that the *ROSAT* source RE0922+710 and the *EUVE* source EUVE J0922+71.1 are separated by only $75''$ on the sky, each within the positional error circle of the other. (The typical 90%-confidence error-circles for *EUVE* and *ROSAT*, while subject to some uncertainties, are of order $80''$.) RE0922+710 was detected only in the 150 Å *ROSAT* band at the $3.9\sigma$ level, whereas EUVE J0922+71.1 was detected only in the 400 Å *EUVE* band at the $2.9\sigma$ level. These sources are not included in the latest compilations of EUV sources by the *ROSAT* and *EUVE* teams (Pye et al. 1995; Lewis et al. 1994; Bowyer et al. 1995) because their significance is slightly below the cutoff adopted for inclusion. They are, however, likely to be real (J. Pye, private communication; X. Wu, private communication).

Maoz et al. (1996) obtained $B$ and $R$ CCD images of the field of these EUV sources, and noticed a blue ($B - R = 0.4$, $R = 18.3$ mag) object lying between the positions of the EUV sources, and within their respective error-circles. A database search showed that the blue object is positionally coincident with a 6 cm radio source (Gregory & Condon 1991) and a faint *ROSAT* PSPC (0.2–2.4 keV) X-ray source (White, Giommi, & Angelini 1994). A finding chart of the field is given in Fig. 1. The blue object, and the positions and error circles of the various detections, are marked. A summary of the positions, count-rates, and estimated fluxes at all wavelengths is given in Table 1.

The positionally-consistent radio through X-ray emission led us to investigate this field further. In particular, we were intrigued by the possibilities that this is an unknown type of Galactic EUV source, or a quasar detected in the EUV through a hole in the Galaxy's interstellar medium (ISM). Note that, apart from the detection of about 20 bright (13–14 mag) active galactic nuclei, all in the 100 Å bands, no extragalactic objects have been detected in the EUV (Lewis et al. 1994; Barber et al. 1995; Marshall, Fruscione, & Carone 1995). EUV radiation is very strongly absorbed by the Galaxy's ISM. We report here optical spectroscopy of the blue object, and a 21 cm H I emission measurement in its direction.

## 2  Optical Spectroscopy

A 5000 – 7600 Å spectrum of the blue object was obtained on 1 February 1995 UT with the Large Cassegrain Spectrograph on the 2.7-m telescope at McDonald Observatory. A Craf-Cassini 1024 × 1024-pixel CCD detector was used with a 300 l/mm grating and a 2″ slit at position angle 90° to give a spectral resolution of less than 10 Å. The total integration time was 2700 s. The spectrum reveals two broad (FWHM 5000–6000 km s$^{-1}$) emission lines at 5300 Å and 6550 Å (C IV $\lambda$1549 and C III] $\lambda$1909), identifying the object as a high-redshift quasar.

Additional CCD spectra were obtained on 24 February 1995 UT with the Kast double spectrograph (Miller & Stone 1993) at the Cassegrain focus of the 3-m Shane reflector at Lick Observatory. The total integration time was 900 s in the range 3120–10400 Å, plus a short (300 s) exposure in the range 4260–7030 Å to verify the reliability of the flux calibration in the dichroic crossover region ($\sim$ 5000–5400 Å). The seeing was about 1″, and thin cirrus may have been present. Although the air mass was relatively high ($\sim 1.7$), the effects of atmospheric dispersion were minimized by orienting the long slit (of width 2″) along the parallactic angle; thus, the relative flux calibration should be accurate. The spectral resolution was $\sim 6$ Å below 5200 Å, and $\sim 14$–17 Å above it.

Fig. 2 shows the Lick spectrum of the quasar. The Ly$\alpha$, C IV, and C III] emission lines are prominent, and indicate a redshift of $z = 2.432$. There is a strong unresolved (FWHM 7Å) absorption line near the peak of the Ly$\alpha$ emission. The spectral slope ($F_\nu \propto \nu^\alpha$) is $\alpha \approx -1.5$, somewhat steeper than that of most quasars (e.g. Richstone & Schmidt 1980). The absolute fluxes of the Lick and McDonald spectra agree to $\sim 10\%$, but are about 25% lower than that implied by the photometric $R$-band measurement (obtained 7 months earlier, on July 1, 1994 UT). This is reasonable considering some seeing losses through the narrow slits, and the possibly non-photometric conditions of the spectroscopic observations. The equivalent widths in the observer's frame of the Ly $\alpha$ + NV, C IV, and C III lines are, respectively, 670 Å, 180 Å, and 100 Å. Their flux ratios are 8.1 : 2.2 : 1.

## 3  21-cm H I Measurements

The identification of the blue object as a quasar raises the possibility that we are seeing a quasar in the EUV through a hole in the ISM. In the H I maps of Stark et al. (1992) the quasar (Galactic coordinates $l^{II} = 141.7°$, $b^{II} = 37.4°$) lies inside several concentric contours of decreasing H I column density. The lowest contour in this "dip" is still fairly high, corresponding to about $2.4 \times 10^{20}$ cm$^{-2}$, but a deeper tunnel may have been smoothed out by the map's resolution ($> 2°$). The lowest total $N_H$ columns that have been measured in the Galaxy are a few $10^{19}$ cm$^{-2}$ (Lockman, Jahoda, & McCammon, 1986), whereas to obtain transparency to light at 200 − 400

Å a column density lower than $\sim 5 \times 10^{18}$ cm$^{-2}$ is required (e.g. Cox & Reynolds 1987). The total count rate in the *ROSAT* PSPC detection is about that expected from an 18-mag quasar, but there are too few (14) total counts to determine the $N_H$ absorbing column by fitting the x-ray spectrum. We therefore carried out a 21 cm measurement of the H I column density in the direction of the quasar.

Observations of Galactic 21 cm H I emission toward 0917+7122 were made with the 43 m NRAO telescope at Green Bank, WV, which has a half-power beam-width of 21' in the 21 cm line. The H I spectra covered about $\pm 250$ km s$^{-1}$ around zero velocity relative to the local standard of rest (LSR). They were corrected for stray 21 cm H I radiation using the technique described by Lockman et al. (1986), although the actual correction amounted to less than 10%. The measured Galactic H I column density toward the quasar is $3.3 \times 10^{20}$ cm$^{-2}$ for an assumed H I spin temperature of 150 K. The estimated 1$\sigma$ uncertainty on the value of $N_H$ is $< 2 \times 10^{19}$ cm$^{-2}$ from all sources, including the opacity correction. The Galactic H I spectrum in a $6° \times 4°$ field around 0917+7122 has two main spectral components, one near zero velocity LSR and the other near $-50$ km s$^{-1}$. The total $N_H$ across this field varies systematically from about $2 \times 10^{20}$ cm$^{-2}$ north of the quasar to about $6 \times 10^{20}$ cm$^{-2}$ in the south because of a change in the zero velocity line strength. Toward 0917+7122 itself, the zero and $-50$ km s$^{-1}$ components have widths of 10 and 20 km s$^{-1}$ (FWHM), respectively, and the negative velocity component contains $7 \times 10^{19}$ cm$^{-2}$. The uniformly large values of $N_H$ in the sky around 0917+7122, the presence of two spectral components, and the fairly large line widths all argue that it is unlikely that there are extreme fluctuations in the total Galactic $N_H$ on angular scales smaller than the 21' beam of the 43 meter telescope in this part of the sky. Thus the present data should give an accurate value of the Galactic $N_H$ toward the quasar.

## 4 Conclusions

As part of a program to search for optical counterparts to unidentified EUV sources, we have optically identified a $z = 2.432$ quasar coincident with a previously detected radio and X-ray source. Although the quasar lies within the error circles of an EUV source detected by both *EUVE* and the *ROSAT*/WFC, 21 cm H I emission measure-ments in this direction show a high total H I column density. It is therefore unlikely that the quasar and the EUV source are associated.

The EUV source is likely to be nearby, within $\sim 100$ pc (Warwick et al. 1994), and only by chance coincidence on the line-of-sight to the quasar. As shown in more detail by Maoz et al. (1996), all the stars within 2.5′ of either of the EUV detections (i.e. well beyond the error circles shown in Fig.1), down to a limit of $R = 22$, $B = 21$, have $B - R$ colors and magnitudes consistent with late-type dwarfs at distances of 200 pc or more. At such distances, these stars are unlikely candidate optical counterparts to the EUV source. This indicates that the EUV source is either very faint optically, or is an unknown type of nearby Galactic star that mimics the color and magnitude of a distant late-type dwarf. In the former case, the EUV/optical flux ratio, $\nu F_\nu(150\text{Å})/\nu F_\nu(6500\text{Å}) > 80$, i.e. the source is some kind of extremely hot object. A third alternative is that the source emits transiently, and was "turned off" at the time of the optical observations. Upcoming observations with the *Hubble Space Telescope* and *EUVE* may provide further clues to its nature.

**Acknowledgements** We thank K. Anderson, N. Craig, A. Laor, H. Netzer, J. Pye, and X. Wu for helpful discussions and suggestions. Astronomy at the Wise Observatory is supported by grants from the Ministry of Science and Technology and from the Israel Academy of Science. The work of A. V. F. and A. J. B. was supported by the National Science Foundation through grant AST–8957063. The University of Texas observations are partially supported by the Space Telescope Science Institute grant HST GO 2578.01-87A RQ-Q. STScI is operated by AURA, Inc., under NASA contract NAS5-26555 The National Radio Astronomy Observatory is operated by Associated Universities, Inc., under a cooperative agreement with the National Science Foundation. This work has made use of the NASA/IPAC Extragalactic Database (NED), which is operated by JPL, Caltech, under contract with NASA.

Table 1: **Positions and Fluxes**

| Band | Instrumental Count Rate | Flux | R.A. (1950) | Dec. (1950) | Positional Error |
|---|---|---|---|---|---|
| 6 cm | – | 92 mJy | 09 : 17 : 49.7 | 71°22′16″ | 12″ |
| $R$ | – | 18.3 mag | 09 : 17 : 49.9 | 71°22′25.5″ | 0.1″ |
| $B$ | – | 18.7 mag | 09 : 17 : 49.9 | 71°22′25.5″ | 0.1″ |
| $EUVE$ 400 Å | 15 counts ks$^{-1}$ | 4.2† | 09 : 18 : 01.3 | 71°22′41″ | ∼ 80″ |
| $ROSAT$ 150 Å | 15 counts ks$^{-1}$ | 1.2† | 09 : 17 : 45.7 | 71°22′40″ | ∼ 80″ |
| $ROSAT$ 0.2–2.4 keV | 2.5 counts ks$^{-1}$ | 5.0† | 09 : 17 : 47.7 | 71°22′15″ | 20″ |

† Units of $10^{-14}$ erg s$^{-1}$ cm$^{-2}$.

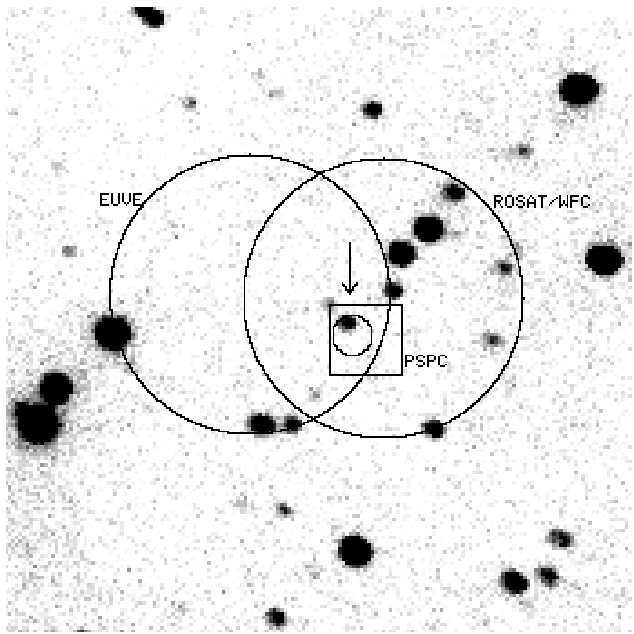

Figure 1: *R*-band CCD image of the field of Q0917+7122. The quasar is marked by an arrow, and the error circles of the various detections are shown. (For clarity, the *ROSAT* PSPC detection is marked with a box rather than a circle.) The small circle marks the 6 cm radio detection. North is up and East is left. The field shown is 6′ square.

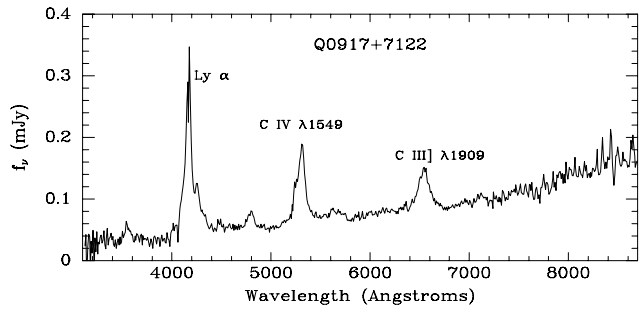

Figure 2: Optical spectrum of Q0917+7122. The narrow features at ∼ 8400 Å are artifacts.

9